\documentclass[prb,notitlepage,twocolumn]{revtex4}

\usepackage{amsmath}
\usepackage{amssymb}
\usepackage{bm}
\usepackage{graphicx}
\usepackage{times}
\begin{document}

\title{Switchable invisibility of dielectric resonators}

\author{Mikhail~V.~Rybin${}^{1,2}$}
\email{m.rybin@mail.ioffe.ru}
\author{Kirill~B.~Samusev${}^{1,2}$}
\author{Polina~V.~Kapitanova${}^{3}$}
\author{Dmitry~S.~Filonov${}^{3}$}
\author{Pavel~A.~Belov${}^{3}$}
\author{Yuri~S.~Kivshar${}^{3,4}$}
\author{Mikhail~F.~Limonov${}^{1,2}$}

\affiliation{$^1$ Ioffe Institute, St.~Petersburg 194021, Russia}
\affiliation{$^2$Department of Dielectric and Semiconductor Photonics, ITMO University, St.Petersburg 197101, Russia}
\affiliation{$^3$Department of Nanophotonics and Metamaterials, ITMO University, St.~Petersburg 197101, Russia}
\affiliation{$^4$Nonlinear Physics Center, Research School of Physics and Engineering,
Australian National University, Canberra ACT 2601, Australia}

\begin{abstract}
The study of invisibility of an infinite dielectric rod with high refractive index is based on the two-dimensional Mie scattering problem, and it suggests strong suppression of scattering due to the Fano interference between spectrally broad nonresonant waves and narrow Mie-resonant modes. However, when the dielectric rod has a finite extension, it becomes a resonator supporting the Fabry-Perot modes which introduce additional scattering and eventually destroy the invisibility. Here we reveal that for shorter rods with modest values of the aspect ratio $r/L$ (where $r$ and $L$ are the radius and length of the rod, respectively), the lowest spectral window of the scattering suppression recovers completely, so that even a finite-size resonator may become invisible. We present a direct experimental verification of the concept of switchable invisibility at microwaves using a cylindrical finite-size resonator with high refractive index.
\end{abstract}

\date{\today}

\maketitle

\section{Introduction}

Invisibility, mimicry, and cloaking are perpetual problems in both wildlife and inorganic nature. A new fresh impetus to the progress in this field has been given by the concept of metamaterials~\cite{fleury2015invisibility}, allowing to design a  cloaking structure that surrounds an object making it invisible to an outside observer. Different approaches to achieve invisibility cloaking have been proposed including multilayered shells that may make even plasmonic particle invisible~\cite{monticone2013full} being also tuned by graphene~\cite{chen2011atomically} or nonlinear layers imbedded into multi-shell structures~\cite{zharova2012nonlinear}, or magneto-optical shell activated by an external magnetic field \cite{kort2013tuning}.  It was  recognized that the invisibility cloaking become possible as a product of the powerful concepts of the so-called transformation  optics~\cite{leonhardt2006optical, pendry2006controlling, schurig2006metamaterial, cai2007optical}, and it was studied for a variety of applications including radar detection, sensors and detectors \cite{alu2009cloaking,alu2010cloaked,fan2012invisible}.


Unfortunately, most of the suggested methods for invisibility cloaking require the creation of specially designed shells or media with perfect parameters making difficult a practical realization of many invisibility concepts \cite{fleury2015invisibility,oliveri2015reconfigurable}. Recently we suggested a novel approach for the realization of tunable invisibility cloaking and a direct switching from visibility to invisibility regimes and back\cite{rybin2015switching}. This approach is based on two basic ideas. 

\begin{figure}[!b]
\includegraphics{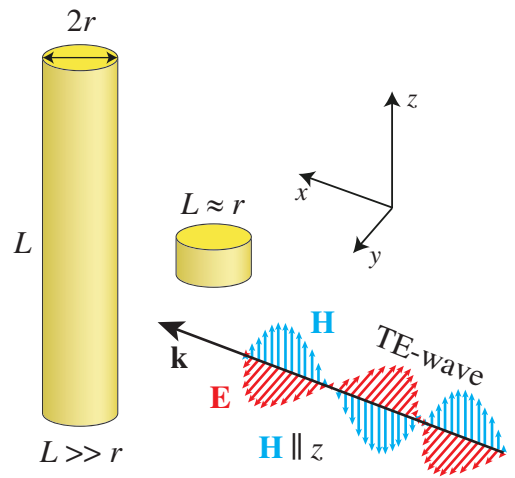}
\caption{
Schematic of the cylindrical dielectric resonator. TE-polarized plane wave incidents to the cylindrical resonator. Aspect $r$-to-$L$ ratio is varied from $r \approx L$ (disk-shaped resonator) to $r \ll L$ (rod-shaped resonator).
}
\label{fig:Scheme}
\end{figure}
%

\begin{figure*}[!t]
\includegraphics[width=12cm]{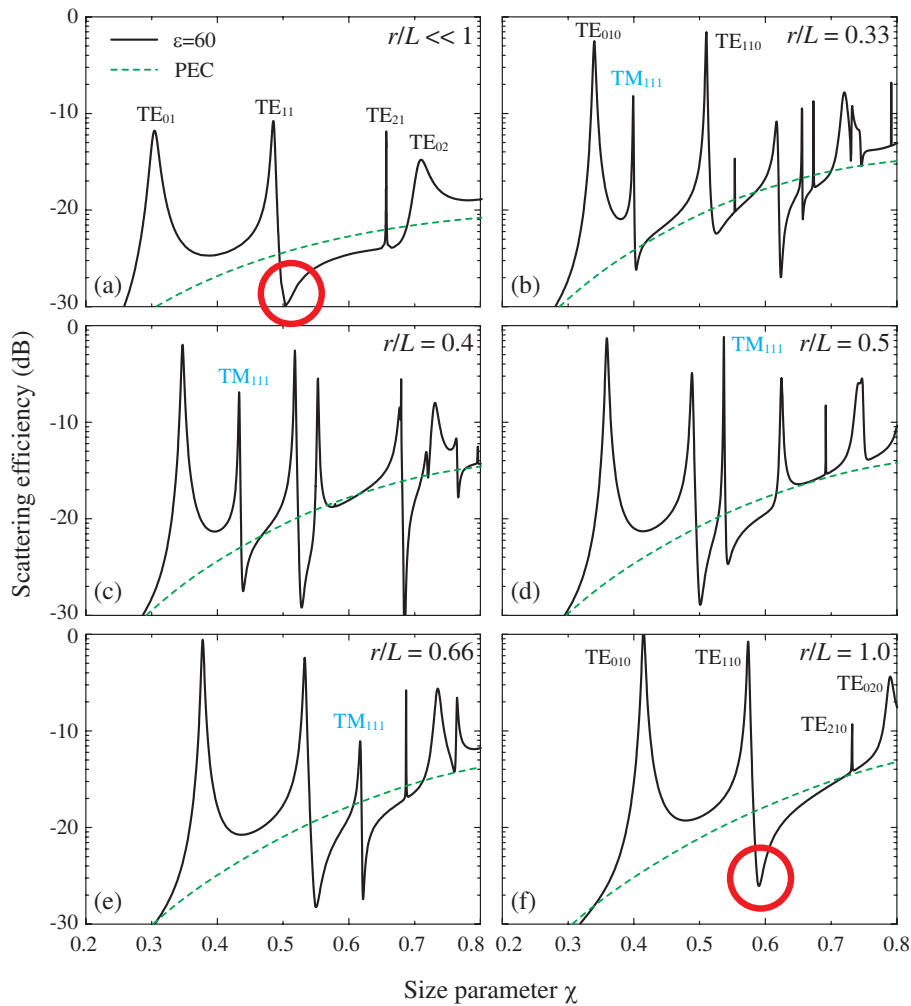}
\caption{ Spectral dependencies of the total scattering efficiency of the infinite homogeneous dielectric cylinder $Q^0$ (a) and finite cylindrical resonator $Q^{r/L} $ for different ratio $r/L$ (b-f) - solid curves. Spectral dependencies of the infinite $Q_{\rm PEC}^{\infty } $ (a) and finite $Q_{\rm PEC}^{r/L} $ (b-f) perfect conducting metal cavities -- dashed curves. TE-polarization, $\varepsilon _{1} =60$, $\varepsilon _{2} =1$. The size parameter $\chi=r\omega /c=2\pi r/\lambda $.}
\label{fig:Qsca}
\end{figure*}
%

The first idea is to archive the scattering cancellation using the properties of the characteristic lineshape of the Fano resonance \cite{ruan2009temporal,chen2012invisibility}. We can expect an interaction of the Fano-type if a spectrally narrow band virtually any origin interacts with broad spectrum (featureless background) through an interference effect constructively or destructively. It is well known that the Fano resonance has been observed across many different branches of physics
\cite{hopfield1967fanoGaP,cerdeira1973fano,Madhavan1998kondo,friedl1990supeconductivity} including plasmonic \cite{lukyanchuk2010fano,stern2014fano,arju2015optical} and dielectric structures \cite{rybin2010bragg,poddubny2012Fano,rybin2013fano,chong2014observation,markovs2015fano}.
If there are no parts of background avoiding the above interaction, the Fano relation takes the form \cite{rybin2015switching}
\begin{equation}
\label{eq:Fano}
I(\omega )=\frac{(q+\Omega )^{2} }{1+\Omega ^{2} } \mathop{\sin }\nolimits^{2} [\Delta (\omega )],
\end{equation}
where $q=\cot \Delta $ is the Fano asymmetry parameter, $\Omega =(\omega -\omega _{0} )/(\Gamma /2)$, where $\omega _{0} $ and $\Gamma $ correspond to the position and the width of the narrow band, $\Delta (\omega )=\varphi _{A} (\omega )-\varphi _{B} (\omega )$ is the phase difference between the narrow resonant and continuum states correspondingly. The main point of the approach is based on the characteristic lineshape of the Fano intensity that vanishes at a specific frequency $\omega _{\rm zero} =\omega _{0} -q\Gamma /2$. It means that the scattering by an object completely vanishes. Note that $\omega _{\rm zero} =\omega _{0} $ at $q=0$, $\omega _{\rm zero} <\omega _{0} $ at $q>0$ and $\omega _{\rm zero} >\omega _{0} $ at $q<0$.

The second idea is to employ the well-known Mie resonances. The case of an infinite cylinder, referred to as the Lorenz-Mie theory, corresponds to a 2D problem of scattering in the plane normal to the symmetry axis $z$, and it involves cylindrical functions in the infinite series of the analytical Mie solution \cite{rybin2013mieOE}. As a result of the interference of the nonresonant and resonant scattering, the spectra of the squared Lorenz-Mie scattering coefficients $\left|a_{n} \right|^{2} $ demonstrate asymmetric profiles with either sharp increase or drop at the resonance frequencies of the cylinder eigenmodes. The scattering coefficients $\left|a_{n} \right|^{2} $ can be presented in the form of infinite cascades of resonances each of which is described by the conventional Fano formula (\ref{eq:Fano}) as was shown for the first time numerically \cite{rybin2015switching,rybin2013mieOE} and then analytically for the cases of infinite cylinder \cite{tribelsky2015giant} and sphere \cite{kong2016fano}.

Employing these two basic ideas, one can demonstrate that the resonant Mie scattering from an infinite dielectric cylinder vanishes at frequencies of $\omega _{\rm zero} $ and the cylinder becomes invisible at any angle of observation. Importantly, for a high-index dielectric cylinder a typical asymmetric Fano profile has a local maximum and local minimum located close to each other. Such a proximity allows to realize {\em switchable invisibility} of a macroscopic object transforming from the visible to invisible regimes and back, without using any coating layers. The idea was confirmed experimentally at microwaves by employing high temperature-dependent dielectric permittivity of heated water in a plastic tube~\cite{rybin2015switching}. In contrast to this earlier study on the scattering and invisibility of an infinite cylinder, here we study the a finite-length cylindrical dielectric resonator (Fig.~\ref{fig:Scheme}) and demonstrate novel features of the invisibility phenomena. 

\section{Scattering from dielectric resonators}

As the starting point, we use the previous results for the elastic scattering by a homogeneous infinite dielectric cylinder of the radius $r$ with dielectric permittivity $\varepsilon _{1} $ embedded in a transparent homogeneous medium with dielectric permittivity of $\varepsilon _{2} $ \cite{rybin2015switching}. For the electromagnetic wave incident perpendicular to the infinite cylinder axis $z$, the two transverse polarizations appear decoupled. As a result, the modes can be classified as either TE modes ($E_{x}$, $E_{y}$, $H_{z}$) with the electric field confined to the ($x$, $y$) plane and the magnetic field polarized along the cylinder axis $z$ or TM modes ($H_{x}$, $H_{y}$, $E_{z}$) for which the magnetic field is oriented perpendicular to the cylinder axis $z$. Here, we consider the TE-polarization, at which the Mie resonances excited in the infinite cylinder are denoted as TE$_{nk} $, ($n$ is integer and $k$ is positive integer) where $n$ is the multipole order, and $k$ is the resonance number. Figure~\ref{fig:Qsca}(a) demonstrates the total Mie scattering efficiency $Q^{0} =\frac{2}{\chi} \sum _{n=-\infty }^{\infty } \left|a_{n} \right|^{2} $ ($\chi=r\omega /c=2\pi r/\lambda $) for the infinite cylinder in the low-frequency part of the spectrum at $\varepsilon _{1} =60$ where the resonant modes TE$_{01} $, TE$_{11} $, TE$_{21} $, and TE$_{02} $ are observed \cite{rybin2013mieOE}. For a reference scattering intensity we choose standard scattering efficiency $Q_{\rm PEC}^{0} $ for the infinite cylinder made from perfect conducting metal (PEC, $\varepsilon _{1} \to \infty $). A window of invisibility with $Q^{0} <Q_{\rm PEC}^{0} $ is clearly seen near the TE$_{11} $ Mie resonance at $\chi\approx 0.5$ [Fig.~\ref{fig:Qsca}(a)].

Next we continue with numerical calculations of the scattering from a finite dielectric cylinder $Q^{r/L} $, namely, a finite-extent dielectric resonator with the radius $r$ and length $L$. The resonator with the dielectric permittivity $\varepsilon _{1}$ embedded in the transparent and homogeneous surrounding medium with the dielectric permittivity of $\varepsilon _{2} =1$. The spectra were calculated by using the CST Microwave Studio software. We calculated the scattering efficiency $Q^{r/L} $ as the radar cross section of the cylindrical resonator divided by the projected cross section of the resonator $S=2rL$. Figures~\ref{fig:Qsca}(b-f) show the total Mie scattering efficiency from finite cylindrical resonator $Q^{r/L} $ for different aspect ratios $r/L$ together with the scattering spectra of infinite perfect conducting metal cavities $Q_{\rm PEC}^{r/L} $. Using the standard nomenclature \cite{zhang2008electromagnetic}, the resonator modes are identified to be TM$_{nkp}$ and TE$_{nkp}$ where $n$ is integer related to the rotational behavior of the mode, $k\geqslant 1$ identifies resonance number for the modes with fixed symmetry $n$, and $p$ enumerates mode relative to the cylinder axis direction. For a rather long cylindrical resonator ($L>r$), the low-frequency spectrum $Q^{r/L} $ demonstrates several strong additional `resonator modes' in comparison with the spectrum of the infinite cylinder $Q^{0} $ (Fig.~\ref{fig:Qsca}). Resonator modes have an obvious origin being related directly to the appearance of the end faces of a finite cylinder.

New (Fabry-Perot or resonator) modes introduce additional scattering that suppress partly or completely spectral windows of the invisibility, specially for the lowest Fano-window corresponding to the TE$_{11} $ Mie resonance. With increasing of $r/L$, all resonances that have a genetic link with the Mie modes of the infinite cylinder demonstrate only a slight high-frequency shift. Alternatively, exactly what one might expect, the resonator modes demonstrate a strong shift from the low to higher frequencies with decreasing of the normalized length $L/r$, as clearly illustrated by the behavior of the TM$_{111} $ resonator mode in Fig.~\ref{fig:Qsca}. As a result, for a resonator with parameters $r/L \ge 1$ the lowest TE$_{110} $ window of invisibility at $\chi\approx 0.6$ completely recovers (marked by red circles in Figs.~\ref{fig:Qsca}a and~\ref{fig:Qsca}f).

\section{Experimental approach}

\begin{figure}[t]
\includegraphics{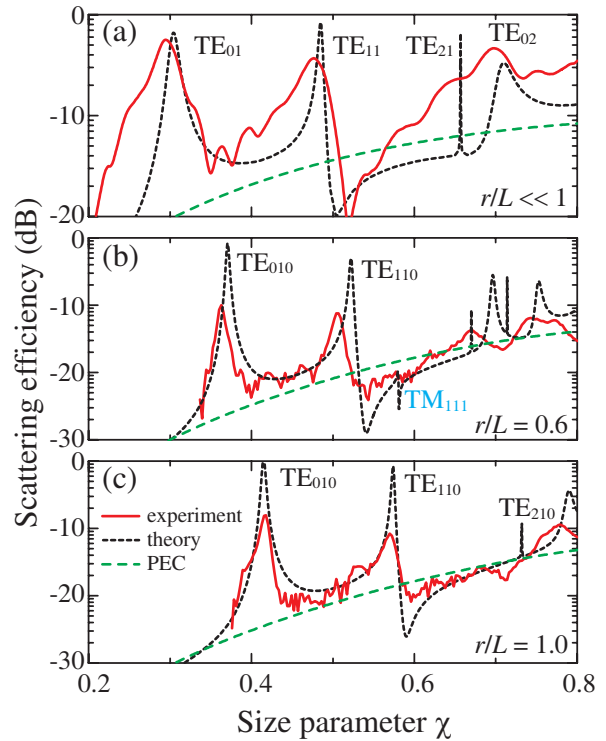}
\caption{ Low-frequency region of the calculated scattering efficiency $Q^{r/L} $ (black short-dashed curve), scattering efficiency of the perfect metallic conductor PEC $Q_{\rm PEC}^{r/L} $ (green long-dashed curve), and the experimentally measured scattering efficiency of a cylindrical tube filled with water (red curve): (a) $r/L = 0$ ($r/L = 0.03$ in experiment), (b) $r/L=0.6$, and (c) $r/L=1$. TE-polarization, $\varepsilon _{1} =60$, $\varepsilon _{2} =1$.}
\label{fig:rh}
\end{figure}
%

\begin{figure*}[t]
\includegraphics{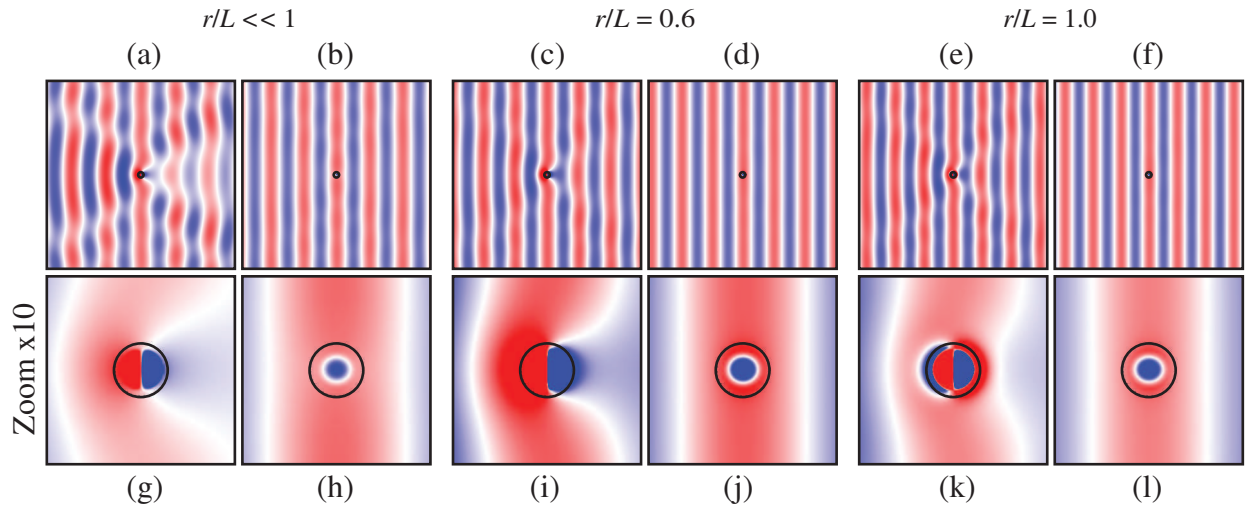}
\caption{Numerical results. Magnetic field map for the $H_{z}$ component around cylinder (a)-(f) and inside cylinder (g)-(l). Shown are: strong Mie scattering (a), (g) for $r/L =0$, $\chi = 0.48$, (c), (i) for $r/L = 0.6$, $\chi = 0.52$, (e, k) for $r/L =1$, $\chi = 0.57$, and the Fano invisibility regime (b), (h) for $r/L=0$, $\chi = 0.50$, (d), (j) for $r/L = 0.6$, $\chi = 0.54$, (f, l) for $r/L =1$, $\chi = 0.59$. TE-polarization, $\varepsilon _{1} =60$, $\varepsilon _{2} =1$.}
\label{fig:FieldMap}
\end{figure*}
%

To demonstrate experimentally the realization of the invisibility of the cylindrical resonator with high refractive index without any coating layers, we use a plastic cylinder filled with water that is characterized by dielectric permittivity of $\varepsilon \approx 80$ at room temperature and $\varepsilon \approx 50$ at 90$^{\circ }$C in the microwave frequency range \cite{ellison2007permittivity}. The microwave scattering spectra $Q_{}^{r/L} $ of the cylindrical resonator depending on the $r/L$ parameter and water temperature were measured in an anechoic chamber. A rectangular horn antenna (TRIM 0.75 GHz to 18 GHz; DR) connected to a transmitting port of a vector network analyzer Agilent E8362C is used to approximate a plane-wave excitation. The cylindrical rod of water with radius $r=20$~mm and variable height is placed into the far-field region of the horn antenna on the distance of 2.5~m. The similar horn antenna is employed as a receiver and placed on 2.5 m distance from the cylindrical rod. For the TE polarization, the electromagnetic wave incident perpendicular to the resonator axis $z$ and the magnetic field is oriented along the axis $z$. Using the optical theorem \cite{g120} we calculate the scattering efficiency from the imaginary part of the measured forward scattering amplitude. The dielectric permittivity of the plastic tube is much smaller than that for water in microwave frequency range and the tube has no effect on the scattering efficiency.

We observe experimentally a strong suppression of scattering at frequencies corresponding to the Fano dips in the function $Q^{r/L} $. Figure~\ref{fig:rh} shows an invisibility effect, namely suppression of scattering lower than PEC level $Q_{\rm PEC}^{r/L} $ due to the TE$_{110}$-Fano resonance at $\chi\approx 0.5$ for $r/L =0$ ($r/L = 0.03$ in experiment) [Fig.~\ref{fig:FieldMap}(a)], at $\chi\approx 0.55$ for $r/L =0.6$ [Fig.~\ref{fig:FieldMap}(b)], and at $\chi\approx 0.6$ for $r/L =1$ [Fig.~\ref{fig:FieldMap}(c)]. Therefore the incident TE-polarized electromagnetic wave passes the cylindrical resonator without scattering making it invisible from any angle of observation. For $r/L =0.6$, the TM$_{111} $ resonator mode at $\chi\approx 0.58$ partly suppress spectral windows of invisibility. Nevertheless, with future decreasing of $r/L$, the TM$_{111} $ resonator mode shifts to higher frequencies and the lowest TE$_{110} $ Fano-window of invisibility completely recovers.

To demonstrate dramatic difference in the scattering between the maximum and minimum of the Fano Mie line-shape, we calculated numerically a structure of the magnetic field around and inside the cylindrical resonator depending on $r/L$. Figure~\ref{fig:FieldMap} shows that for the TE$_{110}$ Mie resonance in the regime of Fano invisibility we observe practically complete suppression of the scattering, see panels (b), (d), and (f). For those normalized frequencies the resonant $H_{z}$ field near the cylinder's boundary inside the rod mimics the incident field distribution, Fig.~\ref{fig:FieldMap}(h),(j),(l). As a result, an incident TE-polarized electromagnetic wave passes the cylindrical resonator without scattering that is the resonator becomes invisible from any angle of observation without additional coating layers. At the same time inside the resonator one can observe very strong Mie resonance which nevertheless do not disturb the incidence wave because of the Fano condition.

\section{Invisibility map}

The calculations show that by adjusting the frequency and dielectric permittivity, a homogeneous high-index cylindrical resonator can be made essentially invisible without any additional cloaking layers.

\begin{figure}[t]
\includegraphics{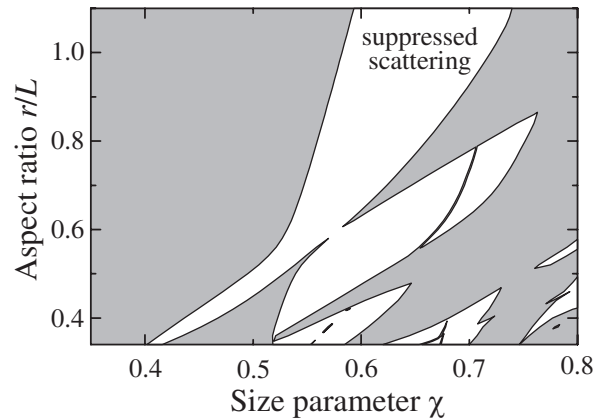}
\caption{Invisibility map for a homogeneous dielectric resonator in the low-frequency spectral region. Shaded areas define the regions of strong scattering. Open domains maps the region of suppressed scattering $Q^{r/L} $ in comparison with the PEC scattering $Q_{\rm PEC}^{r/L} $. TE-polarization, $\varepsilon _{1} =60$, $\varepsilon _{2} =1$.  }
\label{fig:InvisMap}
\end{figure}
%

To analyze the invisibility dynamics, we calculated the difference $Q^{r/L}-Q_{\rm PEC}^{r/L}$ as a function of the resonator dielectric permittivity $\varepsilon _{1}$. We found that the scattering efficiencies of dielectric cylindrical resonator and PEC coincide when $\varepsilon _{1} $ is about 10. At $\varepsilon _{1} >10$ the invisibility dynamics appears for the TE$_{110} $ Mie resonance. The invisibility map demonstrated in Fig.~\ref{fig:InvisMap} shows the regions of invisibility [$Q^{r/L} < Q_{\rm PEC}^{r/L} $] with respect to the controllable parameters $\chi$ and $r/L$ for $\varepsilon _{1} =60$. The regions of the invisibility are shaded in white. From Fig.~\ref{fig:InvisMap} it is clearly seen that the invisibility in low-frequency spectral regions follow strong Fano resonance for TE$_{110} $ Mie mode.

As can be seen from Figs.~\ref{fig:Qsca} and~\ref{fig:InvisMap}, for a rather long cylindrical resonator ($L>r$), the resonator mode TM$_{111} $ introduces additional scattering within TE$_{110} $ window of invisibility. The effect regains with increasing of $r/L$ when the resonator modes demonstrate a strong shift from the low to higher frequencies. For structural parameters $r/L\ge 1$ the lowest window of invisibility completely recovers.

\section{Demonstration of switchable invisibility}

\begin{figure}[t]
\includegraphics{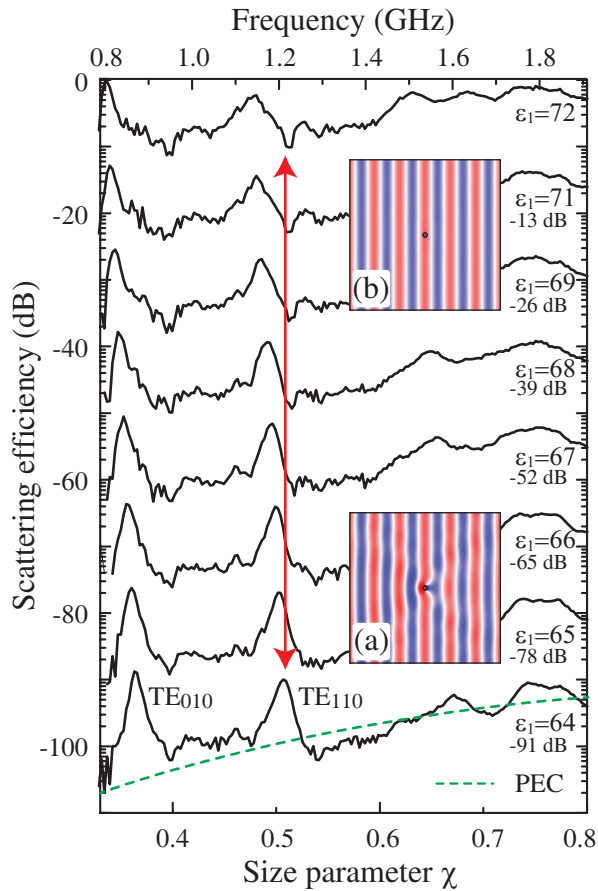}
\caption{Experimental results. (a) Measured temperature dependence of the total scattering efficiency $Q_{}^{r/L} $ of a finite-length ($r/L=0.6$) cylindrical resonator filled with water (TE polarization, $\varepsilon _{2} = 1$). Curves are shifted vertically by the values marked on the plot. Inserts (a, b) show the calculated magnetic fields at the frequency 1.22 GHz in the regimes of the Fano invisibility and strong Mie scattering, respectively.
}
\label{fig:Tuning}
\end{figure}
%

Figure~\ref{fig:Qsca} shows that for the high-index dielectric resonator an asymmetric Fano profile has a maximum and a minimum located close to each other. This proximity can be used for the demonstration of the switching between visibility and invisibility. To prove the concept of Fano invisibility experimentally, we take advantage of the strong temperature dependence of the water permittivity $\varepsilon _{1}$ in the microwave frequency range\cite{ellison2007permittivity}. It changes from $\varepsilon _{1}  = 80$ at 20$^\circ$C to $\varepsilon _{1} $ = 50 at 90${}^\circ$C. Such unique dependence of $\varepsilon _{1} $ leads to a considerable temperature shift of the maxima (strong scattering) and minima (possible invisible regime) of the Fano-Mie resonance modes, as seen in Fig.~\ref{fig:Tuning} for TE$_{010} $ and TE$_{110} $ bands. When high-frequency wing of TE$_{110} $ Fano lineshape moves through a fixed frequency we observe a strong change of the scattering intensity from high to very small values. Therefore, the Fano resonance makes possible to switch the scattering of a cylindrical resonator filed with water from the visible regime of strong Mie scattering ($varepsilon_1=64$, water at $T = 80^\circ$C) to the invisibility regime ($varepsilon_1=72$, water at $T= 35^\circ$C) at the same frequency of 1.22 GHz.

\section{Concluding remarks}

We have demonstrated that different Mie resonances of a homogeneous dielectric resonator with high refractive index are responsible for sharp Fano resonances in the TE-polarized scattered spectra. The unique property of the characteristic Fano lineshape is that the transmission intensity vanishes at a special frequency. This means that the scattering completely vanishes at this frequency, and an object becomes invisible at any angle of observation. The effect of the Fano invisibility has been suggested earlier for an infinite dielectric cylinder in low-frequency spectral region \cite{rybin2015switching}. Here,  we have studied the scattering from a finite-length dielectric resonator and revealed that the spectra of a rather long resonator ($L>r$) contain additional TM$_{nkp}$ and TE$_{nkp}$ ($p \ne 0$) modes that superimpose in the low-frequency region and introduce additional scattering suppressing or completely removing the spectral windows of the invisibility. However, when the ratio $r/L$ becomes larger, the resonator modes shift to higher frequencies and the invisibility window completely recover at $r/L\ge 1$. As a result, a finite-length dielectric resonator becomes invisible from any angle of observation at certain frequencies for the TE-polarized waves. We have verified these theoretical predictions in microwave experiments.

\section*{Acknowledgments}

 We acknowledge fruitful discussions with A.A.~Kaplyanskii. This work was supported by the Russian Foundation for Basic Research (No. 16-02-00461), and the Australian Research Council. The experimental studies have been supported by the grant of Russian Science Foundation (No. 15-19-30023). P.K. acknowledges a scholarship of the President of the Russian Federation.

\end{document}